\documentclass[aps,pra,groupedaddress,twocolumn]{revtex4}
\usepackage{graphicx,amsmath,amssymb}

\begin{document}

\title{Creation of orbital angular momentum states with chiral polaritonic lenses}
\author{Robert Dall$^{1,2}$, Michael D. Fraser$^3$, Anton S. Desyatnikov$^{1}$,  \\ Guangyao Li$^{1}$, Sebastian Brodbeck$^{4}$, Martin Kamp$^{4}$, Christian Schneider$^{4}$,  Sven H\"ofling$^{4,5}$, Elena A. Ostrovskaya$^{1}$}
\affiliation{$^{1}$Nonlinear Physics Centre and $^{2}$AMPL, Research School of Physics and Engineering, The Australian National University, Canberra, ACT 0200, Australia  \\ $^{3}$Quantum Optics Research Group, RIKEN Center for Emergent Matter Science, 2-1 Hirosawa, Wako-shi, Saitama 351-0198, Japan \\ $^4$ Technische Physik and Wilhelm-Conrad-R\"ontgen Research Center for Complex Material Systems, Universit\"at W\"urzburg, D-97074 W\"urzburg, Germany \\ $^5$ School of Physics and Astronomy, St Andrews University, KY16 9SS, United Kingdom}

\date{\today}

\begin{abstract}
Controlled transfer of orbital angular momentum to exciton-polariton Bose-Einstein condensate spontaneously created under incoherent, off-resonant excitation conditions is a long-standing challenge in the field of microcavity polaritonics. We demonstrate, experimentally and theoretically, a simple and efficient approach to generation of nontrivial orbital angular momentum states by using optically-induced potentials -- {\em chiral polaritonic lenses}.
\end{abstract}


\maketitle


{\em Introduction.---} Recent advances in optical excitation and manipulation of exciton-polaritons in semiconductor microcavities lead to creation and trapping of polariton Bose-Einstein condensate \cite{BEC06} in optically induced potentials \cite{pattern_formation, oscillations_nature, ringexp_1,ringexp_2}. These potentials are created by incoherent optical sources of exciton-polaritons due to self-trapping mechanisms that are inherent in this open-dissipative system \cite{gain_guiding,our_pra,isomorphy13} and are similar to those at play in optical systems with gain and loss \cite{rosanov,akhmediev,1d_kartashov_ol}. The advantage of the``soft", optically-induced potentials over those``hard-wired" in the microcavity, e.g., by etching process \cite{etching_06}, is the ability to reconfigure their spatial and energy landscape by structuring the optical pump. 

A long-standing and so far unsolved problem in the exciton-polariton physics is the inability to transfer orbital angular momentum directly from the optical pump to the spontaneously condensed exciton-polaritons. The effective potentials created by optical pump via uncondensed reservoir of high-energy near-excitonic polaritons depend only on pump intensity, and all of the phase information is ``scrambled" in the process of energy relaxation. This is in stark contrast to a condensate of ultracold atoms that admits direct imprinting of quantum states of photons \cite{atom}, and to coherently driven polaritons in the resonant excitation schemes \cite{opo10,opo13}. The solution of this problem holds the key to controlled creation of quantised orbital angular momentum states and persistent currents, which could be employed in the polariton analogue of SQUID sensors  \cite{SQUID_aBEC1,SQUID_aBEC2,SQUID_pBEC} and information encoding devices, as well as in the fundamental studies of vortices \cite{vortex_atom} and polariton Bose-Einstein condensates under rotation. So far, vortices in an incoherently excited exciton-polariton condensate have only been generated spontaneously and, in the absence of total angular momentum in the system, only in the form of vortex-antivortex pairs \cite{pvortex08,Vortex11,vortex11_2,vortex_ring_13} or degenerate spin vortices \cite{Deveaud13,Amo14}.

\begin{figure}[here]
\includegraphics[width=8.5 cm]{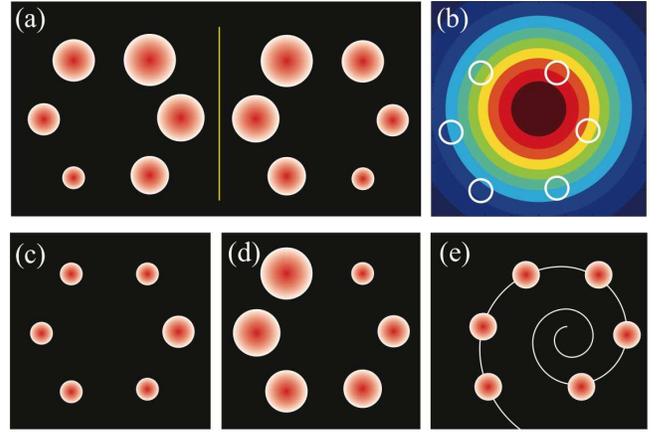} 
\caption{(Color online) Schematics of the optical pump structured via a six-pinhole metal mask. (a) Spatial distribution of intensity shown together with its mirror image to demonstrate chirality: the pump structure cannot be superimposed with its mirror image by simple rotation. Chirality appears due to accidental misalignment as shown in (b); (b) Gaussian pump beam shifted relative to the mask centre, resulting in the spatial intensity distribution shown in (a); (c-e) Engineered intensity distributions of the (c) non-chiral and (d,e) chiral polaritonic lenses tested in experiment. In (a) the intensity of the pump spots is represented by their size, and the degree of asymmetry in (a,c-e) is exaggerated for clarity.}
\label{chirality}
\end{figure}

In this work we show that {\em chiral polaritonic lenses} -- potentials with broken chiral symmetries -- enable reliable creation of nontrivial orbital angular momentum states in spontaneously formed exciton-polariton Bose-Einstein condensates. Such lenses are formed by an optical pump with a {\em chiral distribution of intensity} [Fig. \ref{chirality} (a)] structured with a pinhole optical mask. Chirality of the lens can be {\em accidental} due to small beam shifts relative to the mask [Fig. \ref{chirality} (b)] or {\em engineered} by varying the size or position of pinholes [Fig. \ref{chirality} (d,e)].  In particular, we demonstrate efficient creation of a single charge vortex by a spiral polaritonic lens in Fig. \ref{chirality} (e). This method represents  a new paradigm in control of orbital angular momentum states of light and matter, with all of the schemes previously demonstrated  in optics \cite{OE_vortex13,Natasha14}, plasmonics \cite{SPP}, ultracold atom physics \cite{atom}, and polaritonics \cite{opo10,opo13} relying on coherent, resonant sources with a {\em chiral distribution of phase}.

{\em Experiment.---} In experiment, we work with an GaAs/AlGaAs microcavity sandwiched between distributed Bragg reflector mirrors to achieve high confinement of a photon mode \cite{supp}. In the regime of strong coupling between the photons and excitons confined in 2D quantum wells imbedded in the microcavity, exciton-polaritons form and can be driven to Bose-Einstein condensation. The condensation occurs spontaneously above a threshold power of the pumping laser. 

The {\em cw} optical pump, with the excitation energy far above the exciton or polariton resonance, is a broad Gaussian beam spatially modulated in the near-field by a metal mask patterned with an arbitrary configuration of holes, and subsequently re-imaged onto the microcavity sample surface at normal incidence. It allows us to create an azimuthal distribution of pump spots responsible for trapping of the condensate in the centre of the effective trap produced by the polaritonic lens, similarly to the experiments involving spatial light modulators \cite{ringexp_1,ringexp_2}. The mask enables us to create structured potentials of {\em arbitrary shape} with stability limited only by stability of the laser. For this work, in particular, we use a six-spot azimuthal intensity distribution. 

As the mask is re-imaged on the surface of the sample, the size of the potential is determined by the depth of focus. In the current experiment pump spots of $3$ $\mu$m are separated by $2$ $\mu$m, with the typical size of the resulting condensate $\sim 13$ $\mu m$. The imaging system consists of a free-space microscope with a high numerical aperture objective, which collects the photoluminescence from the sample, and allows us to infer both spatial and momentum distribution of condensed polaritons from emitted photons. The pump is linearly polarized and the resulting photoluminescence appears to be unpolarized. A Michelson interferometer is used to analyse the phase structure of the signal \cite{supp}. 

\begin{figure}[here]
\includegraphics[width=7.0 cm]{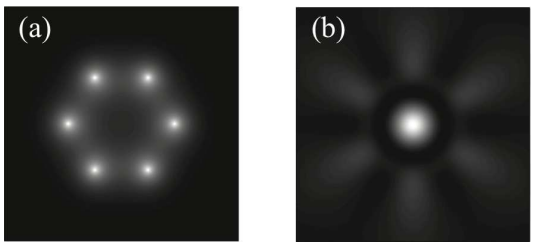} 
\caption{Schematic illustration of linear interference of six phase coherent polariton condensates (a) at low pump intensity, where the fringes are absent due to the long wavelength; (b) at high pump intensity (see text).}\label{linear}
\end{figure}

Control of the polariton flow with a structured optical pump relies on two experimentally verified processes. First, several pump spots in close proximity of each other create phase-locked condensates \cite{oscillations_nature}. And secondly, nonlinear interaction between the polaritons and a reservoir creates an effective trapping potential with the depth proportional to the strength of the pump \cite{ringexp_1,ringexp_2}.  The spatial distribution of the condensate density is affected by the pump intensity. This could be understood from a simple linear interference picture. In the far field of each individual pump spot, in the plane of the quantum well, the polariton matter-wave can be approximated by a wave packet with a radially symmetric phase front and exponentially decaying envelope \cite{our_pra}: $\psi_c\sim\exp(-\Gamma r)\exp(i {\bf k}_p {\bf r})$. As the intensity of the pump increases, so does the chemical potential (energy of the steady state) of the condensate, $\mu$, \cite{Wouters08,our_pra}, and both the wavelength of the polariton matter-wave, $\lambda_p\sim 1/k_p$, and the width of the condensate decrease as $\{\lambda_p,\Gamma\}\sim \mu^{-1/2}$ \cite{our_pra}. For the perfectly symmetric ring-like configuration of six identical pump spots, the superposition of long-wavelength condensate ``tails"  would tend to localise the density in the excitation regions for weaker pump intensity (larger $\lambda_p$ and $\Gamma$) [Fig. \ref{linear}(a)]. A higher pump intensity (smaller $\lambda_p$ and $\Gamma$) would produce or a bright spot in the centre [Fig. \ref{linear}(b)]. In the latter regime, the structured pump works as a lens, focusing the condensate into the centre. These spatial patterns would then be amplified due to stimulated scattering of polaritons into the regions of high density and therefore lead to different spatial structures of steady states at different pump intensities.

\begin{figure}
\vspace{5 mm}
\includegraphics[width=8.5 cm]{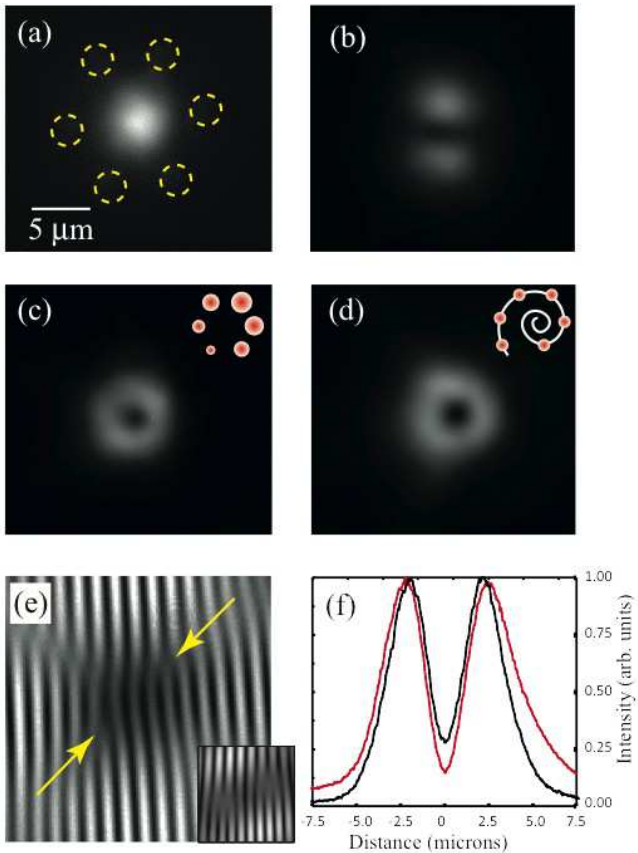} 
\caption{(Color online) Experimental real space image of cavity emission intensity showing (a) ground state  ($|m|=0$), (b) dipole mode ($|m|=0$), and (c,d) charge one vortices ($|m|=1$) created by the polaritonic lenses in Figs. \ref{chirality}(a,e). Circles in (a) mark positions of pump spots; (e) Experimental and (inset) theoretical interference pattern of retroreflected far-field photoluminescence from the charge one vortex in (d). Arrows mark locations of ``forks" indicating presence of a single isolated vortex in the centre of the condensate. (d) Intensity profiles of the vortices in (c,d) obtained with accidentally (black line) and deliberately engineered (red line) chiral lens at the pump intensities marked by dashed lines in Fig. \ref{population} (a,b). Images in (a-d) and (f) are plotted on the same scale, and the interferometric image (e) is magnified for clarity.}
\label{interference}
\end{figure}

The condensate formed in the centre of the polaritonic lens just above the condensation threshold is almost perfectly radially symmetric [Fig. \ref{interference} (a)], and can be thought of as a ground state of a radially symmetric 2D potential well effectively created by the pump through induced spatial distribution of the reservoir density \cite{ringexp_1,ringexp_2}. In the polar coordinates, the eigenstates of this potential can be written as $\psi_{n,m}=\Phi_n(r)\exp(i m\theta)$. The first excited energy state produced by a stronger pump is a {\em dipole mode} superposition [Fig. \ref{interference} (b)] of two $n=1$  states with the nonzero orbital angular  momentum, i.e. quantised vortices with topological charge $m=+1$ and $m=-1$. The second excited state is a {\em quadrupole mode} superposition (not shown) of $n=2$ vortex states with the topological charge $m=+2$ and $m=-2$ and a state with $m=0$ containing a radial node. The total topological charge and orbital angular momentum of the superposition states is zero. To select a mode with non-zero topological charge, one needs to break the chiral symmetry of the lens.

Even a slight misalignment of the Gaussian pump beam with the centre of the metal mask used to re-image sophisticated spatial distribution of intensity onto the surface of the sample can create symmetry breaking in the polaritonic lens [Fig. \ref{chirality} (c,d)]. As shown in Fig. \ref{chirality}, the resulting structures can be non-chiral (c) or chiral (a,d). At a low pump power, the {\em accidental chirality} is weak, and the resulting higher-order state observed in the experiment still resembles a dipole mode [Fig. \ref{interference} (b)]. However, as the pump intensity grows, the mode selection strongly favours a single charge one vortex, $m=\pm1$, clearly visible both in real space images [Fig. \ref{interference} (c)] and interferometric images [Fig. \ref{interference} (e)] of the photoluminescence signal. Thus, {\em accidental chirality forces the polariton condensate into a nonÐzero orbital angular momentum state}.

\begin{figure}
\includegraphics[width=8.5 cm]{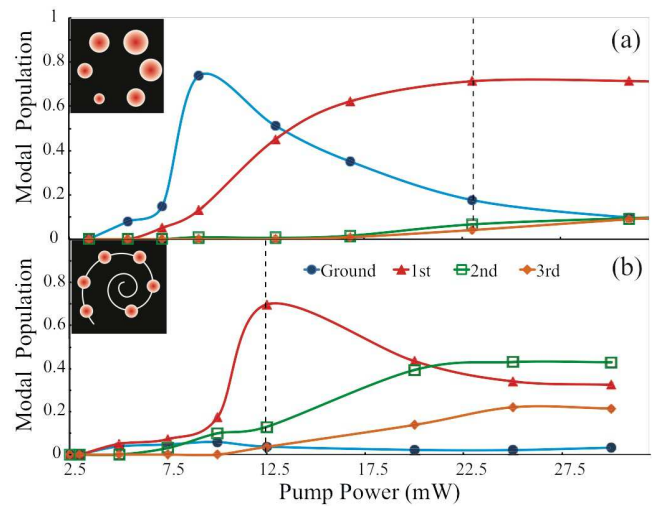} 
\caption{(Color online) Experimental measurement of relative populations of the ground ($n=0$), first ($n=1$), and higher-order ($n=2,3$) energy states of the effective trap induced by the polaritonic lens with (a) accidental and (b) engineered chirality. The condensation threshold in both cases is $P_{\rm th}\approx 5$ mW. Solid lines are a guide to the eye. Dashed lines mark vortex states ($|m|=1$) shown in Fig. \ref{interference}(c,d). }
\label{population}
\end{figure}

The mode selection of a vortex state is much more efficient in polaritonic lenses with {\em engineered chirality}. In the experiment, we tested both the circular masks with different hole sizes [as shown in Fig. \ref{chirality} (d)] and the spiral masks with identical holes [Fig. \ref{chirality} (e)]. The efficiency of the mode selection is quantified via the energy resolved measurement of the mode population vs. pump power. As seen in Fig. \ref{population}(b), the engineered chiral structure {\em strongly suppresses} formation of the ground state condensate with zero orbital angular momentum \cite{supp}. The ground state mode clearly visible in Fig. \ref{population}(a) at lower powers, e.g., at $\sim 10$ mW,  is very weakly populated in Fig. \ref{population}(b). The vortex $m=\pm1$ state produced by the engineered chiral lens has a stronger admixture of the $m=\pm2$ state, which results in a much greater contrast, defined through the minimum and maximum intensity of the photoluminescence as $I_c=(I_{max}-I_{min})/I_{max}$. The admixture of the ground state causes the contrast to deviate from $1$ (for a perfect zero intensity in the vortex core). For the vortex produced with the engineered chiral structure in our experiments [Fig. \ref{interference} (d)]  the best contrast is $I_c \approx 0.82$ [at $P=12.5$ mW, dashed line in Fig. \ref{population}(b)], whereas the best contrast achieved for accidental chirality is $I_c \approx 0.76$ [at $P=22.5$ mW, dashed line in Fig. \ref{population}(a)]. We note that the efficient mode selection by the {\em incoherent} polaritonic lens represents more than a two-fold improvement on the contrast $I_c\approx 0.38$ demonstrated in the {\em coherent} vortex excitation experiments via the resonant OPO scheme \cite{opo10}. 

\begin{figure}[here]
\includegraphics[width=8.5 cm]{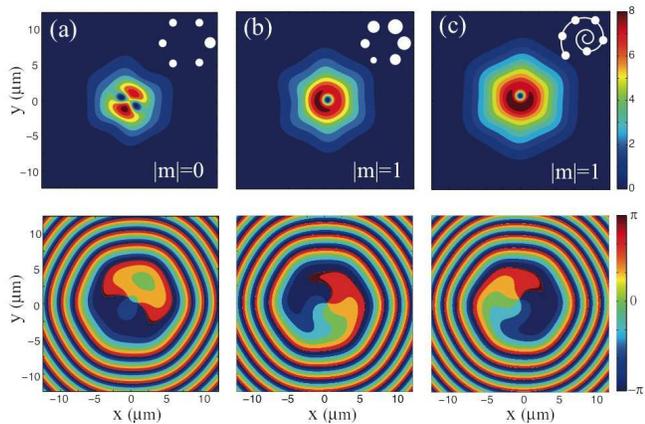} 
\caption{(Color online) Modelled steady state real-space density (top row) and phase (bottom row) created with (a) the non-chiral hexagonal polaritonic lens shown in Fig. \ref{chirality}(c) for the unbalanced pump with $P/P_{\rm th}=1.17$; (b) polaritonic lens with {\em accidental chirality}  with $P/P_{\rm th}=1.2$;  (c) the {\em engineered chiral lens}. Insets show schematic configuration of the polaritonic lenses, and $|m|$ indicates total topological charge of the resulting structure. }
\label{unbalanced}
\end{figure}

{\em Modeling.---} We consider a spontaneously formed exciton-polariton condensate under the continuous-wave far off-resonant optical excitation injecting free carriers into the system high above the lower polariton energy. The model consists of a mean-field equation for the polariton condensate wavefunction and a rate equation for the inhomogeneous density of the reservoir \cite{Wouters07}:  
\begin{eqnarray} 
i\hbar\frac{\partial \Psi}{\partial t}&=&\left[-\frac{\hbar^2}{2m}\nabla_\perp^2+V(\vec{r},t)+ i\frac{\hbar}{2}(Rn_R-\gamma_c)\right]\Psi,  \nonumber \\  
\frac{\partial n_R}{\partial t}&=&-(\gamma_R+R|\Psi|^2)n_R(\vec{r},t)+P(\vec{r}), \label{model} 
\end{eqnarray}
where $V(\vec{r},t)=g_c|\Psi|^2+g_Rn_R(\vec{r},t)$. Here $\Psi$ is the condensate wavefunction, $n_R$ is the reservoir density, and $P(\vec{r})$ is the spatially modulated optical pump. The critical parameters defining the condensate dynamics are the loss rates of the polaritons $\gamma_{c}$ and reservoir polaritons $\gamma_{R}$, the stimulated scattering rate $R$, and the strengths of polariton-polariton, $g_c$, and polariton-reservoir, $g_R$, interactions. In what follows, we use the dimensionless form of the model obtained by using the scaling units of time, energy, and length: $T=1/\gamma_c$, $E=\hbar \gamma_c$,  $L=[\hbar/({m_{LP}\gamma_c})]^{1/2}$, where $m_{LP}$ is the lower-polariton effective mass \cite{parameters}.

%\begin{figure}
%\includegraphics[width=8.5 cm]{chiral_vortices} 
%\caption{(Color online) Modelled polariton distributions within the chiral hexagonal polaritonic lenses. (a) Density and (b) phase distribution of a stable charge one vortex obtained in the chiral lens shown in Fig. \ref{chirality}(a).  (c) Density and (d) phase distribution of a stable charge two vortex obtained in the chiral lens shown in Fig. \ref{chirality}(d). The pump intensity varies by $\delta P=1\%$ from spot to spot, $\delta P_{max}=5 \%$, $P/P_{th}=1.2$ Insets show schematic configurations of the polaritonic lenses.}
%\label{vortices}
%\end{figure}

We model both chiral and non-chiral polaritonic lenses with very small asymmetries, so that the spot-to-spot variation of intensity similar to those shown in Fig. \ref{chirality}(a) is $1\% \geq \delta P \leq 5\%$. First, we model the weakly unbalanced {\em non-chiral} lens shown in Fig. \ref{chirality}(c). These are produced, e.g., by beam shifts along a symmetry axis of the mask. For the pump intensity at threshold and no symmetry breaking ($\delta P=0$), formation of ground state condensate with peak density in the geometrical centre was seen in numerical calculations and in our experiments, and was also observed in recent experiments \cite{ringexp_1} for similar polaritonic lenses of small radii. In numerical simulations, the threshold for condensation is roughly determined by the ratio $P_{\rm th}=\gamma_R\gamma_c/R$ \cite{Wouters07}.

Even the weak symmetry breaking results in a drastically different polariton density distribution. Above the threshold, the condensate favours formation of a steady state with the {\em dipole-mode} structure [Fig. \ref{unbalanced}(a)]. In simulations, the two lobes of the dipole are separated by a phase fold binding a stable vortex-antivortex pair. Such pairs may form spontaneously  due to nonlinearity-induced instabilities \cite{lev} and are destroyed as they move to the periphery of the condensate, unless a special density profile is engineered to hold them in place \cite{vortex_ring_13}. The remarkable feature here is the survival of a single stable vortex-antivortex pair and the resulting formation of the stationary dipole mode. We stress that, since the polaritonic lens is non-chiral, the system does not distinguish between left- and right-hand circulation of polariton flows. Thus, the symmetry breaking of this kind cannot lead to generation of a single isolated vortex in the system.

Next, we model the lenses with accidental [Fig. \ref{chirality}(a,b)] or engineered [Fig. \ref{chirality} (d,e)] chirality. The drastic consequence of the introduced handedness of the system is the formation and stabilisation of  steady states containing {\em single vortices} [Fig. \ref{unbalanced}(b,c)]. We stress that the vortices appear strictly due to the symmetry breaking in the chiral polaritonic lens and, in numerical modelling, no vorticity was ``seeded" into the system. In principle, vortices of various topological charge corresponding to the different degree of asymmetry in the system can be generated. The detailed study of this process  will be reported elsewhere.

We also note that the model (\ref{model}) does capture the experimentally observed resemblance of the collective polariton modes to eigenstates of a linear potential well created in the middle of a polaritonic lens. Indeed, according to the model, the effective linear potential created for polaritons $V_{\rm lin}\sim (g_R\gamma_c/R){\bar P}({\bf r})$, where ${\bar P}=P/P_{\rm th}$, takes form of the strongly repulsive (anti-trapping) barrier localised around the periphery of the polaritonic lens, thus creating a trap in the middle. The self-induced nonlinear contribution to the potential due to polariton interactions is, to the leading order, $V_{\rm nl}\sim g_c[1-{\bar P}({\bf r})(g_R\gamma_c)/(g_c\gamma_R)]|\Psi|^2$, and therefore acts to enhance the trapping potential for the chosen parameters, which agrees well with previous studies \cite{ringexp_1,ringexp_2} and our experimental observations.

{\em Conclusions.---} We have demonstrated operation of chiral polaritonic lenses for creation and trapping of incoherently excited polariton condensates with non-zero orbital angular momentum, containing single vortices. The role of the broken symmetry in such lenses is two-fold: first, in the presence of strong polariton interactions, it triggers the development of nonlinear instabilities leading to formation of phase vortices and anti-vortices, and secondly, the distinct handedness of the system leads to selection of steady states with isolated single vortices and overall non-zero orbital angular momentum.  Our findings open the way to construction of all-optical elements for shaping and directing of polariton flows with a well-defined orbital angular momentum, which could be captured by  potentials ``hard-wired" in a microcavity and used for study of vortices and persistent currents.

It is tempting to draw parallels between the optical manipulation of exciton-polaritons in the plane of the quantum well and shaping of radiation by means of surface plasmon-polariton lenses \cite{SPP}. However, as stressed in the introduction, shaping of the optical wavefront  in the coherently illuminated nano-structures relies on precise {\em spiral phase distribution} introduced by the asymmetrically placed {\em coherent} sources. In contrast, formation of vortex states in an {\em incoherently} excited exciton-polariton condensate is a result of the mode selection in an effective potential induced by a {\em spiral intensity distribution}, and therefore is much less sensitive to the precise geometry of the structured pump. Although exciton-polaritons allow to manipulate light on the micro-scale rather than a nanoscale, the ultrafast velocities and strong nonlinearities inherent to exciton-polaritons and unavailable in plasmonics could potentially enable novel optoelectronic devices. 

{\em Acknowledgments.---} This work was supported by the Australian Research Council (ARC) and State of Bavaria. We are indebted to K. Y. Bliokh for careful reading of the manuscript and in-depth comments. Discussions with I. V. Shadrivov and Yu. S. Kivshar are gratefully acknowledged.


\begin{thebibliography}{99}
\bibitem{parameters} We choose the physical parameters close to those determined for our experimental sample, i.e., $m_{LP}=10^{-4}m_e$, $g_c=6\times 10^{-3}$ meV$\mu$m$^2$, $g_R=2g_c$, $\gamma_c=0.1$ ps$^{-1}$, $\gamma_r=3.5 \gamma_c$, and $R=0.01$ $\mu$m$^2$ps$^{-1}$. Here $m_e$ is the electron mass in vacuum.
\end{thebibliography}
\end{document}